\documentclass[twocolumn,fleqn,A4paper]{articlex}

\DeclareMathAlphabet\mathcal{OMS}{cmsy}{m}{n}
\DeclareMathAlphabet\mathbfcal{OMS}{cmsy}{b}{n}

\usepackage{t1enc,epsfig,latexsym} 
\usepackage[T1]{fontenc}
\usepackage{lmodern}
\usepackage{subfigure}
\usepackage{amsmath}
\usepackage{graphicx}
\usepackage{multirow}
\usepackage{rotating}
\usepackage{sistyle}
\usepackage[superscript]{cite}
\usepackage{lipsum}
\usepackage{mathptmx}
\usepackage{geometry}
\usepackage{verbatim}
\usepackage{hyperref}

\usepackage{bm}
\usepackage{mathtools}
\usepackage{amssymb}
\usepackage{booktabs}
\usepackage{balance}
\usepackage{epstopdf, epsfig}
\usepackage{nicefrac}
\usepackage{savesym}
\savesymbol{nomenclature}
\usepackage{nomencl}
\restoresymbol{OWN}{nomenclature}

\usepackage{etoolbox}
\patchcmd{\thebibliography}
  {\advance\leftmargin\labelsep}
  {\leftmargin=0pt\itemindent=\labelwidth\advance\itemindent\labelsep}
  {}{}

\newcommand{\Eq}[1]{Eq. (\ref{eq:#1})}

\newcommand{\Fig}[1]{Fig. (\ref{fig:#1})}

\newcommand{\Tab}[1]{Table (\ref{tab:#1})}

\newcommand{\bVec}[1]{\bm{#1}}
\DeclareMathAlphabet\mathbfcal{OMS}{cmsy}{b}{n}

\newcommand{\Dbar}[0]{\text{\emph{\DJ}}}

\newcommand{\etal}[0]{ et al.}

\makeatletter

\def\thenomenclature{%
  \@ifundefined{chapter}%
  {
    \section*{\nomname}
    \if@intoc\addcontentsline{toc}{section}{\nomname}\fi%
  }%
  {
    \chapter*{\nomname}
    \if@intoc\addcontentsline{toc}{chapter}{\nomname}\fi%
  }%

  \nompreamble
  \list{}{%
    \labelwidth\nom@tempdim
    \leftmargin\labelwidth
    \advance\leftmargin\labelsep
    \itemsep\nomitemsep
    \let\makelabel\nomlabel}
    }

\RequirePackage{ifthen}
   \renewcommand{\nomgroup}[1]{%
     \ifthenelse{\equal{#1}{G}}{}{%
     	\ifthenelse{\equal{#1}{L}}{}{%
         \ifthenelse{\equal{#1}{S}}{\item[]\item[\textit{Sub/superscripts}]}{}}}}

\usepackage{titlesec}
\titleformat*{\section}{\scshape}
\titlespacing*{\section}{0pt}{12pt}{0pt}

\usepackage{soulutf8}
\titleformat{\subsection}{\normalfont}{\bfserie\thesubsection}{0em}{\ul}

\renewcommand{\figurename}{Figure}
\usepackage[labelsep=period,belowskip=-13pt,aboveskip=7pt,justification=centering,font=large]{caption}

\title{\fontsize{14}{1em}\selectfont On the Rheology of Newtonian Single-Phase Multicomponent Mixtures} 
\author{Sverre Gullikstad Johnsen$^{1,2}$\\ \\ 
$^{1}$SINTEF Industry, Trondheim, Norway\\
$^{2}$NTNU, dept. Materials Science and Engineering, Trondheim, Norway}
\date{} 
\begin{document} 
\maketitle 
\par\vspace*{-6.1cm}
\leftline{ \fontfamily{helvet}\selectfont    }
\par\vspace*{5.2cm}
\renewcommand{\thepage}{\large\sf\arabic{page}}
\begin{large}

\vspace{-12pt}
\section*{\scalebox{.935}[1.0]{ABSTRACT}}
In the turbulent boundary layer of multicomponent fluid mixtures, the species-specific mass flux is determined by the combination of turbulent-diffusiophoretic diffusion and diffusion due to gradients in supplementary fields (e.g. temperature).
For inert mixtures, a balance must exist between all the diffusive transport mechanisms so that the net diffusive mass flux normal to the wall is zero everywhere.
This may require non-constant composition profiles.

Implications are discussed, and mathematical modelling is employed to demonstrate how this may affect fluid property profiles, wall heat flux, and wall shear stress in a Newtonian ternary gas mixture ($H_2 + N_2 + CO_2$) subject to a temperature gradient.

\section*{\scalebox{.935}[1.0]{INTRODUCTION}}
In their classical experiment, Duncan and Toor\cite{Duncan1962} showed that the diffusive transport in an ideal ternary gas mixture of hydrogen ($H_2$), nitrogen ($N_2$), and carbon dioxide ($CO_2$) could not be described satisfactorily by the Fickian formulation.
E.g., the observed development in local nitrogen concentrations could only be explained mathematically by allowing uphill diffusion.
The Duncan-Toor experiments have been further investigated and discussed by e.g. Taylor and Krishna\cite{Taylor93} and Krishna and Wesselingh \cite{Krishna1997}.
It has been shown that Maxwell-Stefan diffusion predicts the non-Fickian behavior observed by Duncan and Toor, accurately.

Whereas the Duncan-Toor experiments were performed under isothermal conditions, Bogatyrev \etal\cite{Bogatyrev2015} studied thermophoresis in binary, ternary, and quaternary mixtures including the ternary $H_2-N_2-CO_2$ mixture.
They emphasized that thermophoresis in multicomponent mixtures depends on the mixture composition in a complex way.

In this paper, it is hypothesized that non-Fickian behavior can cause non-constant composition profiles in the turbulent boundary layer of inert mixtures.
Zero net transport for each species is required, in the direction normal to the wall, but it is suggested that competing diffusive processes (e.g. turbulent, diffusiophoretic and thermophoretic diffusion) which cancel each other out can occur.
This implies that non-constant mass-fraction profiles may be necessary to give zero net diffusive transport.

The concern is that these spatial composition variations will affect fluid properties (e.g. mass density, viscosity, heat capacity, and thermal conductivity) hence the wall heat flux and wall shear stress.
Thus, without the proper understanding, interpretation of  rheology measurements may fail to give a correct assessment of the fluid properties, even for relatively simple  Newtonian ideal mixtures.

Using the ideal ternary gas mixture of Duncan and Toor\cite{Duncan1962} as an example, mathematical modelling of the species transport in the fully developed turbulent boundary layer is employed to support the hypothesis.
Comparing simulations with and without diffusion, it is demonstrated that a significant effect on wall heat flux and wall shear stress can be expected from the diffusion-induced non-constant composition profiles.

\section*{\scalebox{.935}[1.0]{MATHEMATICAL MODELS}}
We are considering a single-phase fluid mixture consisting of a set of $N$ unique, distinguishable, inert species. 
It is assumed that each species field, hence the fluid itself, can be modeled as a continuum. 
This implies that species properties are well defined, continuously varying physical fields throughout the fluid domain. 
Furthermore, it is assumed homogeneous mixing in the sense that local species properties are taken as volume averages over infinitesimal volumes. 
These assumptions allow the utilization of differential calculus in deriving governing equations for the species transport.

\subsection*{Governing Equations}\label{sec:gov_eq}
The set of steady-state governing equations consists of the Advection-Diffusion equation (ADE) for each species,
\begin{equation}
	\bVec{\nabla}\left(\rho_fX_i\bVec{u}_f\right) + \bVec{\nabla}\bVec{j}_{d,i} = 0~,
\end{equation}
the fluid mixture momentum and energy equations,
\begin{equation}
	\bVec{\nabla}\left(\rho_f\bVec{u}_f\bVec{u}_f\right) = -\bVec{\nabla}P + \bVec{\nabla}\bVec{\tau} + \rho_f\bVec{g}~,
\end{equation}
\begin{multline}
	\bVec{\nabla}\left(\rho_fh_{sens,f}\bVec{u}_f\right) =\\
	 \bVec{\nabla}\left(k_f\bVec{\nabla}T\right) - \bVec{\nabla}\left(\sum_{i=1}^N{\bVec{j}_{d,i}h_{sens,i}}\right)~,
\end{multline}
and the restriction that the mass- and mole-fractions must sum to unity,
\begin{equation}\label{eq:sumtounity}
	\sum_{i=1}^N{X_i}=\sum_{i=1}^N{z_i}=1~.
\end{equation}
Introducing turbulence, dimensionless variables (see Appendix) and appropriate simplifications, the simplified governing equations are obtained:
\begin{equation}\label{eq:SimpADE}
	\left(\nicefrac{\nu _t^+\rho_f^+}{Sc_t}\right)\partial_\bot X_i - j_{d,i,\bot}^+=0~,
\end{equation}
gives the mass-fraction profiles; 
\begin{equation}
	\partial_\bot u_{f,\|}^+ = \nicefrac{1}{\left(\mu^+ + \mu _t^+\right)}~,
\end{equation}
gives the dimensionless axial fluid mixture velocity profile; and 
\begin{align}\label{eq:SimpEnEq}
	\partial_\bot \left[k_t^+ \left(\partial_\bot \ln c_P^+\right) T^+ + \left(k_f^++k_t^+\right)\partial_\bot T^+ \right]=0~,
\end{align}
gives the dimensionless temperature profile. 
The $\bot$ and $\|$ indicate the directions normal to and parallel with the wall and the bulk flow direction, respectively, and $\partial_\bot$ denotes the dimensionless gradient component in the direction perpendicular to the wall.
Fore more details, refer to Johnsen \emph{et al.}\cite{Johnsen15}.

\subsection*{Diffusion flux}\label{sec:diff_flux}
Employing Maxwell-Stefan theory \cite{Taylor93}, the dimensionless diffusive mass flux of species $i$ normal to the wall can be expressed as
\begin{align}
	&j_{d,i,\bot }^+=-\rho_f^+D_{ij}^+\partial_\bot\mu_j\label{eq:diffmassflux1}\\
	&\quad~~=-\rho_f^+ D_{ij}^+\left[\Gamma_{jk}\Lambda_{kl}\partial_\bot X_l + d_{\psi,j}\partial_\bot \psi\right]~,\label{eq:diffmassflux2}
\end{align}
where Einstein summation is employed, and the diffusive driving force consists of two terms; namely a diffusiophoretic term due to composition gradients, and a phoretic term due to gradients in other scalar fields (e.g. temperature).
The $D_{ij}$ are the multicomponent diffusion coefficients, 
${\Gamma }_{jk} = \nicefrac{\partial_{z_k} \mu_j}{RT}$, ${{\Lambda }_{kl}}{{\partial }_{\bot }}{{X}_{l}}={{\partial }_{\bot }}{{z}_{k}}$, $d_{\psi,j}=\nicefrac{\partial_\psi\mu_j}{RT}$. 
The chemical potential of species $j$ is expressed as $\mu_j=\mu^0_j+\mu^\psi_j+RT\ln\left(\gamma_jz_j\right)$, where $\mu^\psi_j$ represents the potential contribution from the supplementary, scalar fields.
In the presence of a temperature gradient, the supplementary field gradient can be written
\begin{equation}
	d_{\psi,j}\partial_\bot\psi = \left(\nicefrac{\partial_T\mu_j}{R}\right)\partial_\bot\ln{\left(T^++T_{wall}^{0+}\right)}~.
\end{equation}

It follows from the definitions, that the diffusive mass flux of the $N$th, dependent species is given by $j_{d,N,\bot}^+=-\sum\nolimits_{i=1}^{N-1}{j_{d,i,\bot}^+}$. 
Hence it suffices to solve Eqs. (\ref{eq:SimpADE})-(\ref{eq:SimpEnEq}) for $i\in\left\{1,\dots,N-1\right\}$.

Combining Eqs. (\ref{eq:SimpADE}) and (\ref{eq:diffmassflux1}), it is seen that zero net mass transport can only be ensured by requiring that
\begin{equation}
	D_{ij}^+\partial_\bot\mu_j = -\left(\nicefrac{\nu_t^+}{Sc_t}\right)\partial_\bot X_i~.
\end{equation}
In the following two paragraphs, the implications of this requirement is investigated for two scenarios: 1) the absence of supplementary field gradients ($\partial_\bot\psi=0$); and 2) the presence of a temperature gradient ($\partial_\bot\psi=\partial_\bot\ln{\left(T^++T_{wall}^{0+}\right)}$).

\subsection*{Turbulent-Diffusiophoretic Diffusion}\label{sec:isothermal}
In the case of $d_{\psi,j}\partial_\bot\psi=0$, \Eq{SimpADE} can be written as the homogeneous system of equations
\begin{equation}\label{eq:isothermalADE}
	\mathcal D_{X,il}^+\partial_\bot X_l=0~,
\end{equation}
where $\mathcal D_{X,il}^+ = \left[\left(\nicefrac{\nu _t^+}{Sc_t}\right)\delta_{il} + D_{ij}^+\Gamma_{jk}\Lambda_{kl}\right]$, and $\delta_{il}$ is the Kronecker delta.
It is readily shown that \Eq{isothermalADE} has a non-trivial solution ($\partial_\bot X_l\neq0$) if and only if $-\nicefrac{\nu_t^+}{Sc_t}$ is an eigenvalue of the matrix product $\bVec{D}^+\bVec{\Gamma}\bVec{\Lambda}$.

At the wall, where $\nicefrac{\nu_t^+}{Sc_t}\to0$, the required condition for non-trivial solution reduces to $\det{\left(\bVec{\Gamma}\right)}=0$, since both $\bVec{D}^+$ and $\bVec{\Lambda}$ are invertible.
For ideal mixtures, $\gamma_j=1$ for all $j$, so 
\begin{equation}\label{eq:GammaIdeal}
	\Gamma_{jk,\text{ideal}}=\nicefrac{\delta_{jk}}{z_j}~.
\end{equation}
Hence, ideal mixtures permit the trivial solution ($\partial_\bot X_l=0$) only, at the wall, absent supplementary field gradients.

\subsection*{Combined Turbulent-Diffusiophoretic and Thermophoretic Diffusion}\label{sec:nonisothermal}
In the presence of thermophoresis (due to temperature gradients), there must be a balance between the turbulent-diffusiophoretic diffusion on one side and thermophoretic diffusion on the other.
This can be expressed as the nonhomogeneous system of equations
\begin{equation}\label{eq:nonisothermalADE}
	\mathcal D_{X,il}^+\partial_\bot X_l=-\mathcal D_{T,i}^+\partial_\bot \ln \left(T^+ + T_{wall}^{0+}\right)~,
\end{equation}
where $\mathcal D_{T,i}^+=D_{ij}^+d_{T,j}$.
It is evident that non-zero mass-fraction gradients are required to counter the thermophoresis, in general.

At the wall, \Eq{nonisothermalADE} reduces to
\begin{equation}\label{eq:idealwallcond}
	\left.\Gamma_{jk}\partial_\bot z_k\right|_{wall} = -d_{T,j,wall}\nicefrac{Pr_{wall}}{T_{wall}^{0+}} ~.
\end{equation}
For ideal mixtures, Eqs. (\ref{eq:sumtounity}) and (\ref{eq:idealwallcond}) require that $\sum_{j=1}^N{z_jd_{T,j}}=0$ holds at the wall.
This implies that $d_{T}$s of both positive and negative values must exist at the wall, for ideal mixtures.



\section*{\scalebox{.935}[1.0]{MODEL FLUID}}
The model fluid is a ternary, calorically perfect mixture of perfect gasses consisting of $50$, $25$, and $25~\text{mass-\%}$ of $H_2$, $N_2$, and $CO_2$, respectively.
Species specific heat capacities were extracted from the NIST Chemistry WebBook\cite{NISTChem} while species specific viscosities and thermal conductivities were calculated based on Lennard-Jones parameters found in Andersson\cite{Anderson06}.
Details regarding the modelling of species and mixture material properties (mass density, viscosity, etc.) can be found in Johnsen \etal\cite{Johnsen15}.
Species specific input data are summarized in \Tab{MatProp}.

\begin{table}[!ht]
\vspace{0.5cm}
  \centering
  \caption{Species specific properties.}
    \begin{tabular}{lllll}
          &       &       & \multicolumn{2}{c}{Lennard-Jones param.} \\
          & \multicolumn{1}{c}{$M_w$} & \multicolumn{1}{c}{$c_P^1$} & \multicolumn{1}{c}{$d$} & \multicolumn{1}{c}{$\Omega^1$} \\
          & \multicolumn{1}{c}{$[\nicefrac{\mathrm{kg}}{\mathrm{mol}}]$} & \multicolumn{1}{c}{$[\nicefrac{\mathrm{J}}{\mathrm{molK}}]$} & \multicolumn{1}{c}{$[\mathrm{\text{\AA}}]$} & \multicolumn{1}{c}{$[-]$} \\
    \midrule
    $H_2$ & \multicolumn{1}{l}{0.002016} & \multicolumn{1}{c}{28.84} & \multicolumn{1}{c}{2.915} & \multicolumn{1}{c}{0.857} \\
    $N_2$ & \multicolumn{1}{l}{0.02801} & \multicolumn{1}{c}{29.12} & \multicolumn{1}{c}{3.681} & \multicolumn{1}{c}{1.022} \\
    $CO_2$ & \multicolumn{1}{l}{0.04401} & \multicolumn{1}{c}{37.12} & \multicolumn{1}{c}{3.996} & \multicolumn{1}{c}{1.296} \\
    \midrule
    \multicolumn{5}{l}{$^1$ values at $298\mathrm{K}.$} \\
    \end{tabular}%
  \label{tab:MatProp}%
\end{table}%

For a ternary mixture, there are two independent mass-fraction equations in addition to the velocity and temperature equations.
Moreover, there are only two independent diffusive mass-fluxes, and the matrices that take part in \Eq{diffmassflux2} are $2\times2$ matrices.

The elements of the diffusivity matrix, $\bVec{D}$, can be expressed as\cite{Taylor93}
\begin{equation}
\begin{aligned}
	D_{11}&={\Dbar_{13}\left[ {{z}_{1}}{{{\Dbar}}_{23}}+\left( 1-{{z}_{1}} \right){{{\Dbar}}_{12}} \right]}/{S}~,\\ 
  {{{{D}}}_{12}}&={{{z}_{1}}{{{\Dbar}}_{23}}\left[ {{{\Dbar}}_{13}}-{{{\Dbar}}_{12}} \right]}/{S}~,\\ 
  {{{{D}}}_{21}}&={{{z}_{2}}{{{\Dbar}}_{13}}\left[ {{{\Dbar}}_{23}}-{{{\Dbar}}_{12}} \right]}/{S}~,\\ 
  {{{{D}}}_{22}}&={{{{\Dbar}}_{23}}\left[ {{z}_{2}}{{{\Dbar}}_{13}}+\left( 1-{{z}_{2}} \right){{{\Dbar}}_{12}} \right]}/{S}~,
  \end{aligned}
\end{equation}
where $S={{z}_{1}}{{\Dbar}_{23}}+{{z}_{2}}{{\Dbar}_{13}}+{{z}_{3}}{{\Dbar}_{12}}$, the $\Dbar_{ij}$ are the binary Maxwell-Stefan diffusion coefficients, and $1,2$ and $3$ relate to $H_2$, $N_2$, and $CO_2$, respectively. 
The binary Maxwell-Stefan diffusivities employed by Duncan and Toor\cite{Duncan1962} are cited in \Tab{Dbarval}.
It is noted that the Onsager reciprocal relation implies that $\Dbar_{ij}=\Dbar_{ji}$\cite{Hirschfelder64,Muckenfuss73}.

\begin{table}[!ht]
	\vspace{0.5cm}
  \centering
  \caption{Binary Maxwell-Stefan diffusion coefficients for the ternary $H_2-N_2-CO_2$ mixture\cite{Duncan1962}.}
    \begin{tabular}{ll}
    \toprule
    $H_2 - N_2$ & $\Dbar_{12}=8.33\cdot10^{-5}\nicefrac{\mathrm{m^2}}{\mathrm{s}}$ \\
    $H_2 - CO_2$ & $\Dbar_{13}=6.80\cdot10^{-5}\nicefrac{\mathrm{m^2}}{\mathrm{s}}$ \\
    $N_2 - CO_2$ & $\Dbar_{23}=1.68\cdot10^{-5}\nicefrac{\mathrm{m^2}}{\mathrm{s}}$ \\
    \bottomrule
    \end{tabular}%
  \label{tab:Dbarval}%
\end{table}%

Bogatyrev \etal\cite{Bogatyrev2015} reported thermal diffusion factors, $\alpha_{T}$, as functions of composition for each of the mixture species.
The thermal diffusion factors are related to the diffuiophoretic driving force, $d_{T,j}$ via the thermal diffusion ratio, $k_{T,k}$, by\cite{vandervalk63}
\begin{equation}\label{eq:dTj}
	d_{T,j} = \Gamma_{jk}k_{T,k}~,
\end{equation}
where
\begin{equation}
	k_{T,k} = z_k\sum_{\substack{l=1 \\ l\neq k}}^N{z_l\alpha_{T,kl}}~.
\end{equation}
For ideal mixtures, \Eq{dTj} reduces to
\begin{equation}\label{eq:dTjideal}
	d_{T,j} = \sum_{\substack{l=1 \\ l\neq j}}^N{z_l\alpha_{T,jl}}~.
\end{equation}
Using the experimental data points at $z_l\approx0.5$ from Bogatyrev \etal\cite{Bogatyrev2015}, the thermal diffusion factors and thermophoretic driving force coefficients given in \Tab{thermaldiffratio} were obtained.
For simplicity, constant $d_{T,j}$ were employed in the simulations.

\begin{table}[!ht]
	\vspace{0.5cm}
  	\caption{Thermal diffusion factors, $\alpha_{T,jl}$, based on data from Bogatyrev \etal\cite{Bogatyrev2015} and resulting thermophoretic driving force coefficients, $d_{T,j}$ (assuming ideal mixture, see \Eq{dTjideal}).}
  \centering
	\begin{tabular}{lcccc}
	\boldmath{}\textbf{$j$}\unboldmath{} & \boldmath{}\textbf{$\alpha_{T,j,H_2}$}\unboldmath{} & \boldmath{}\textbf{$\alpha_{T,j,N_2}$}\unboldmath{} & \boldmath{}\textbf{$\alpha_{T,j,CO_2}$}\unboldmath{} & \boldmath{}\textbf{$d_{T,j}$}\unboldmath{} \\
	\midrule
	$H_2$ 	&       & 0.32  & 0.38  & 0.161 \\
	$N_2$ 	& 0.24  &       & 0.06  & 0.073 \\
	\bottomrule
	\end{tabular}%
  \label{tab:thermaldiffratio}
\end{table}%

\section*{\scalebox{.935}[1.0]{SIMULATION SETUP}}
The equations were solved in a numerical modelling framework described by Johnsen \etal\cite{Johnsen15}.
The simulations assume fully developed turbulent flow parallel to the wall.
Moreover, it is assumed that gradients in the main flow direction are negligible and that gradients perpendicular to the wall vanish in the bulk.
Additional details can be found in Johnsen \etal\cite{Johnsen15}.

The wall and bulk temperatures were set equal to the Bogatyrev \etal\cite{Bogatyrev2015} temperatures of $280K$ and $800K$, respectively, and a range of bulk flow velocities were employed.
The boundary conditions employed in the simulations are summarized in \Tab{BC}.

The simulations were conducted on a 1-dimensional computational mesh consisting of $30$ grid points logarithmically distributed between the wall and the bulk.
A grid sensitivity study showed that the wall heat flux and wall shear stress varied with less than 1\% between a grid with 30 grid points and one with 100 grid points.
The first grid point was located $10^{-7} \mathrm{m}$ away from the wall, and the bulk node was located $10^{-3}\mathrm{m}$ away from the wall.
The results were insensitive to decreasing the first node distance to the wall.

To isolate the effect of non-zero composition gradients, simulations with and without diffusion were conducted.
In the simulations without diffusion, the multicomponent diffusion coefficients were zero, $D_{ij}=0~\forall~i,j$.

\begin{table}[!ht]
	\vspace{0.5cm}
  \centering
  \caption{Boundary Conditions employed in simulations.}
    \begin{tabular}{llll}
    \textbf{Boundary Condition} & \textbf{Variable} & \textbf{Value} & \textbf{Unit} \\
    \midrule
    bulk mass-fractions & $X_{H_2,bulk}$ & 0.5   & $\nicefrac{\mathrm{kg}}{\mathrm{kg}}$ \\
          & $X_{N_2,bulk}$ & 0.25  & $\nicefrac{\mathrm{kg}}{\mathrm{kg}}$ \\
    wall diffusion mass flux & $j_{d,H_2,\bot,wall}$ & 0     & $\nicefrac{\mathrm{kg}}{\mathrm{m^2s}}$ \\
          & $j_{d,N_2,\bot,wall}$ & 0     & $\nicefrac{\mathrm{kg}}{\mathrm{m^2s}}$ \\
    bulk temperature & $T_{bulk}$ & 800   & $\mathrm{K}$ \\
    wall temperature & $T_{wall}$ & 280   & $\mathrm{K}$ \\
    bulk flow velocity & $u_{x,bulk}$ & 1,2,5,10 & $\nicefrac{\mathrm{m}}{\mathrm{s}}$ \\
    \bottomrule
    \end{tabular}%
  \label{tab:BC}%
\end{table}%

\section*{\scalebox{.935}[1.0]{SIMULATION RESULTS}}
Simulations were performed with and without diffusion.
In the simulations without diffusion, the mass-fraction profiles were constant throughout the boundary layer, and fluid properties varied only due to the varying temperature.
In simulations including diffusion, however, non-constant mass-fraction profiles resulted to balance the thermophoretic diffusion by turbulent-diffusiophoretic diffusion, to maintain zero net diffusive transport.
The resulting mass-fraction profiles are shown in \Fig{MassFrac}, for the various bulk flow velocities (darker curve corresponds to higher velocity).
Generally, the mass-fraction of $CO_2$ increased towards the wall while $H_2$ and $N_2$ mass-fractions decreased.
Due to the composition dependency in fluid properties (e.g. mass density and viscosity), the simulations predict a bulk flow velocity dependency in these.

\begin{figure}[!tb]
	\begin{center} 
	\hspace{-0.03cm}
	\includegraphics[width=\columnwidth, angle=0]{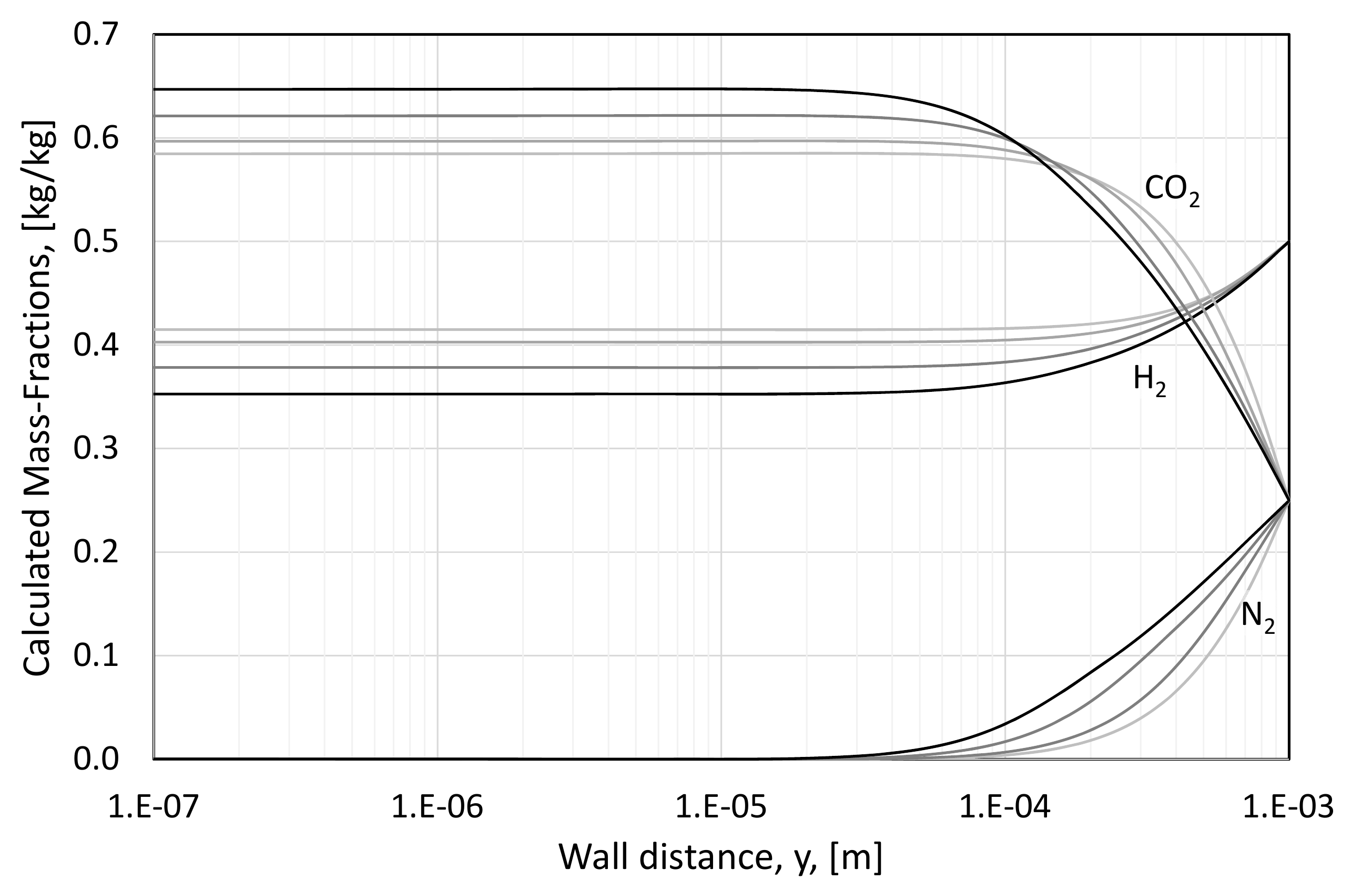}
	\caption{Calculated mass-fractions plotted against wall distance, for the three species $H_2$, $N_2$, and $CO_2$, for the bulk flow velocities $1$ (light grey), $2$, $5$, and $10\nicefrac{\mathrm{m}}{\mathrm{s}}$ (black).}
	\label{fig:MassFrac}  
	\end{center}
\end{figure}

In Figs. (\ref{fig:DenWall}) and (\ref{fig:ViscWall}), respectively, the wall mass density and viscosity are shown as functions of the bulk flow velocity.
It is seen that the effect of diffusion is to reduce the mass density and increase the viscosity.
In the absence of diffusion, the mass density and viscosity are insensitive to the flow velocity since the wall temperature was fixed.

\begin{figure}[!tb]
	\begin{center} 
	\hspace{-0.03cm}
	\includegraphics[width=\columnwidth, angle=0]{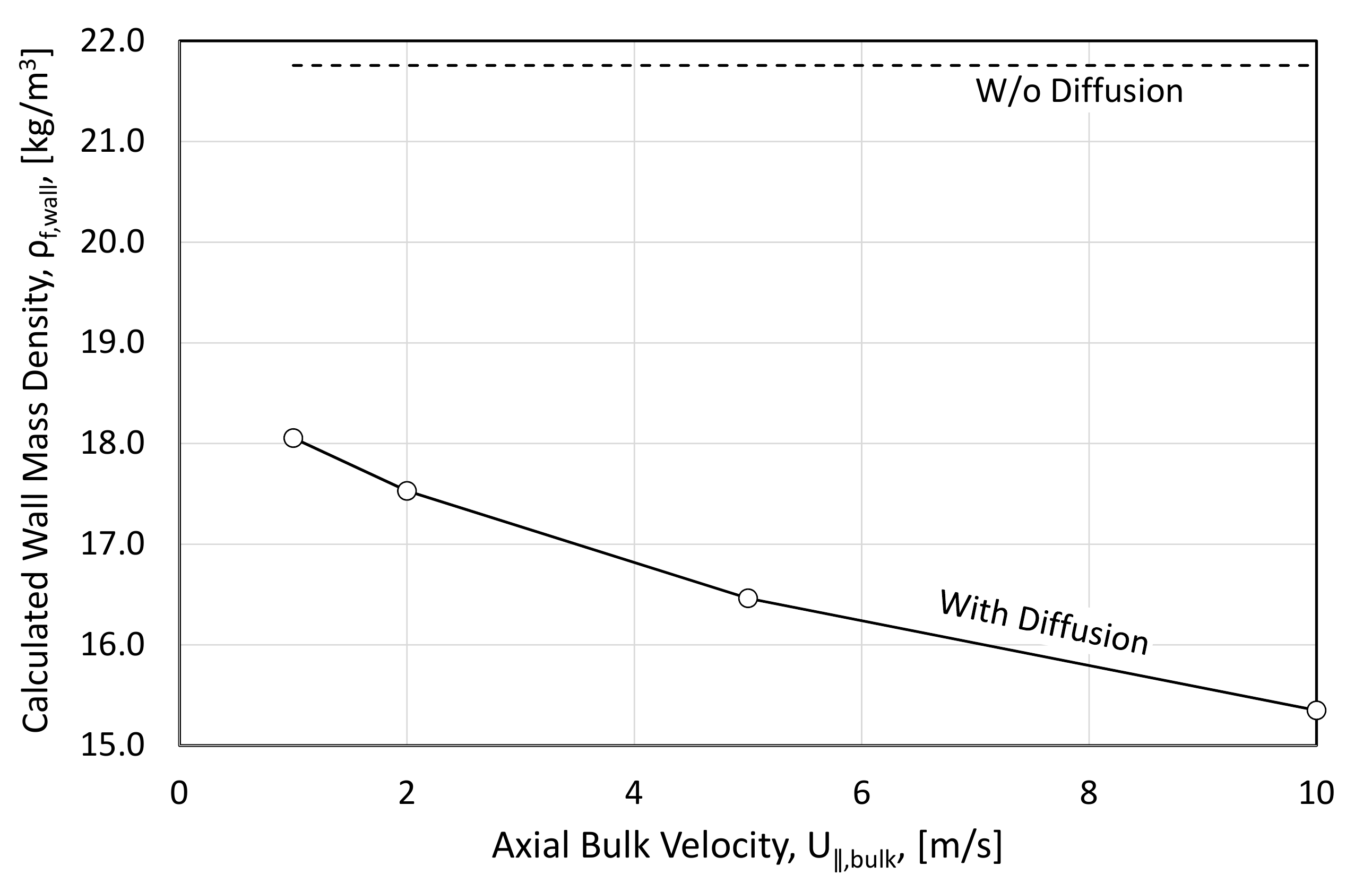}
	\caption{Calculated wall mass density, $\rho_{f,wall}$, plotted against bulk flow velocity.}
	\label{fig:DenWall}  
	\end{center}
\end{figure}
\begin{figure}[!tb]
	\begin{center} 
	\hspace{-0.03cm}
	\includegraphics[width=\columnwidth, angle=0]{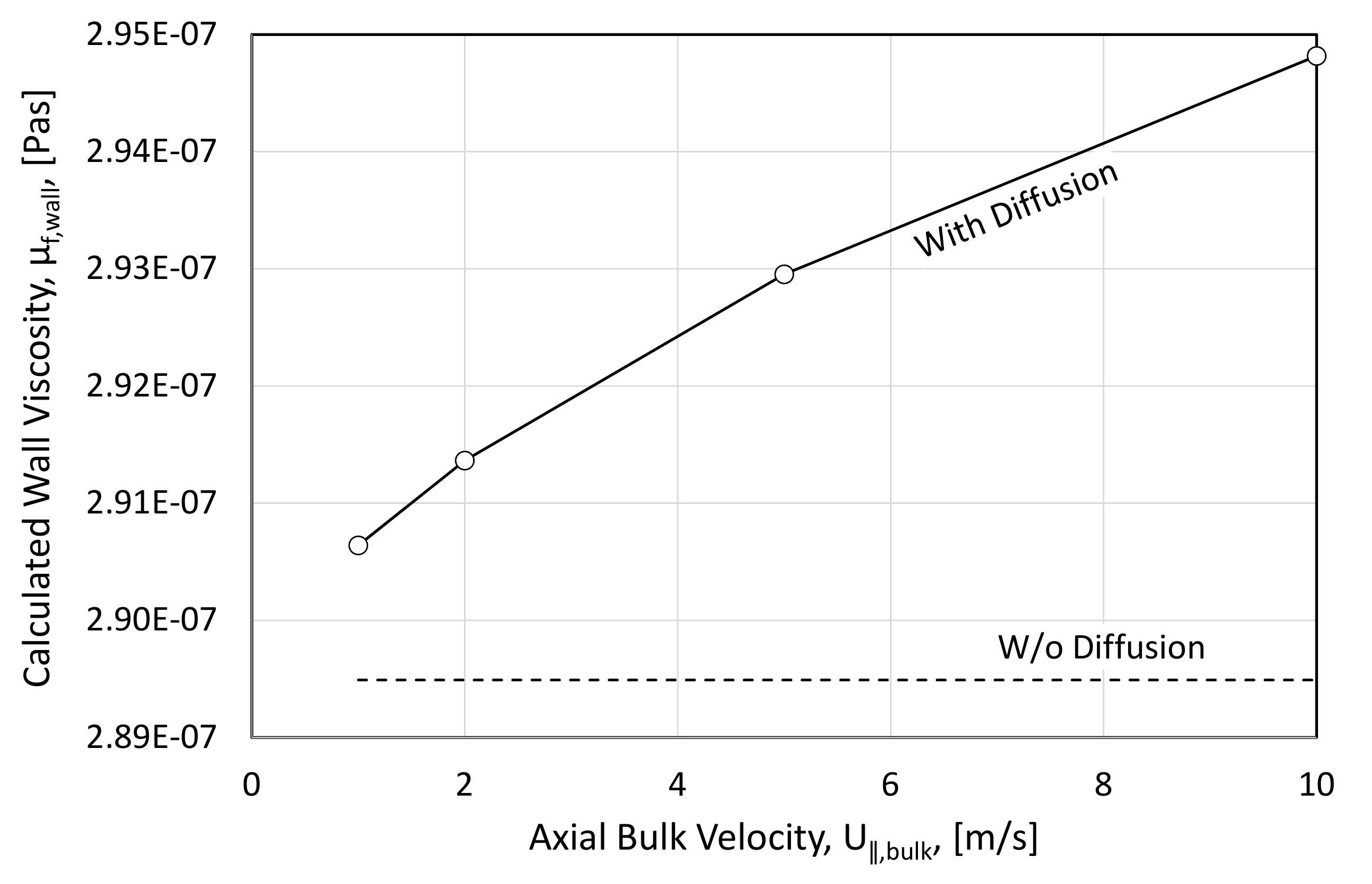}
	\caption{Calculated wall viscosity, $\mu_{f,wall}$, plotted against bulk flow velocity.}
	\label{fig:ViscWall}  
	\end{center}
\end{figure}
\begin{figure}[!tb]
	\begin{center} 
	\hspace{-0.03cm}
	\includegraphics[width=\columnwidth, angle=0]{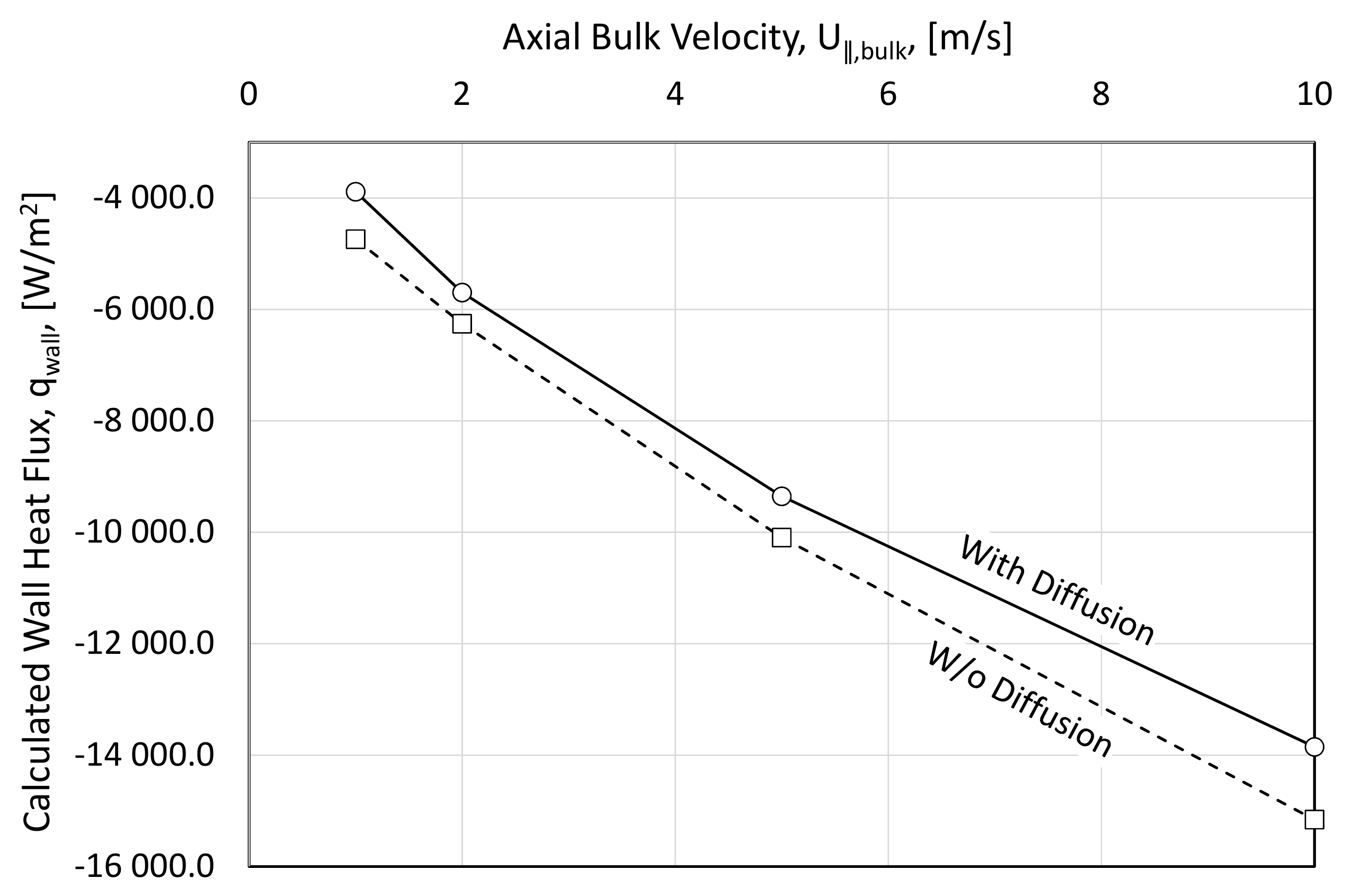}
	\caption{Calculated wall heat flux, $q_{wall}$, plotted against bulk flow velocity (negative heat flux indicates that heat is flowing from the fluid into the wall).}
	\label{fig:QWall}  
	\end{center}
\end{figure}

In \Fig{QWall}, the wall heat fluxes are shown as functions of bulk flow velocity, for simulations with and without diffusion.
Negative heat flux indicates that the heat flows from the fluid into the wall, and the magnitude of the heat flux generally increases with the bulk flow velocity, as expected.
The simulations predict that diffusion will reduce the efficiency of the heat exchange between the bulk and wall.

In \Fig{TauWall}, the wall shear stresses are shown as functions of the bulk flow velocity, for simulations with and without diffusion.
The wall shear stress increases with increasing flow velocity, as expected, but the simulations predict that diffusion will reduce the growth rate.

\begin{figure}[!tb]
	\begin{center} 
	\hspace{-0.03cm}
	\includegraphics[width=\columnwidth, angle=0]{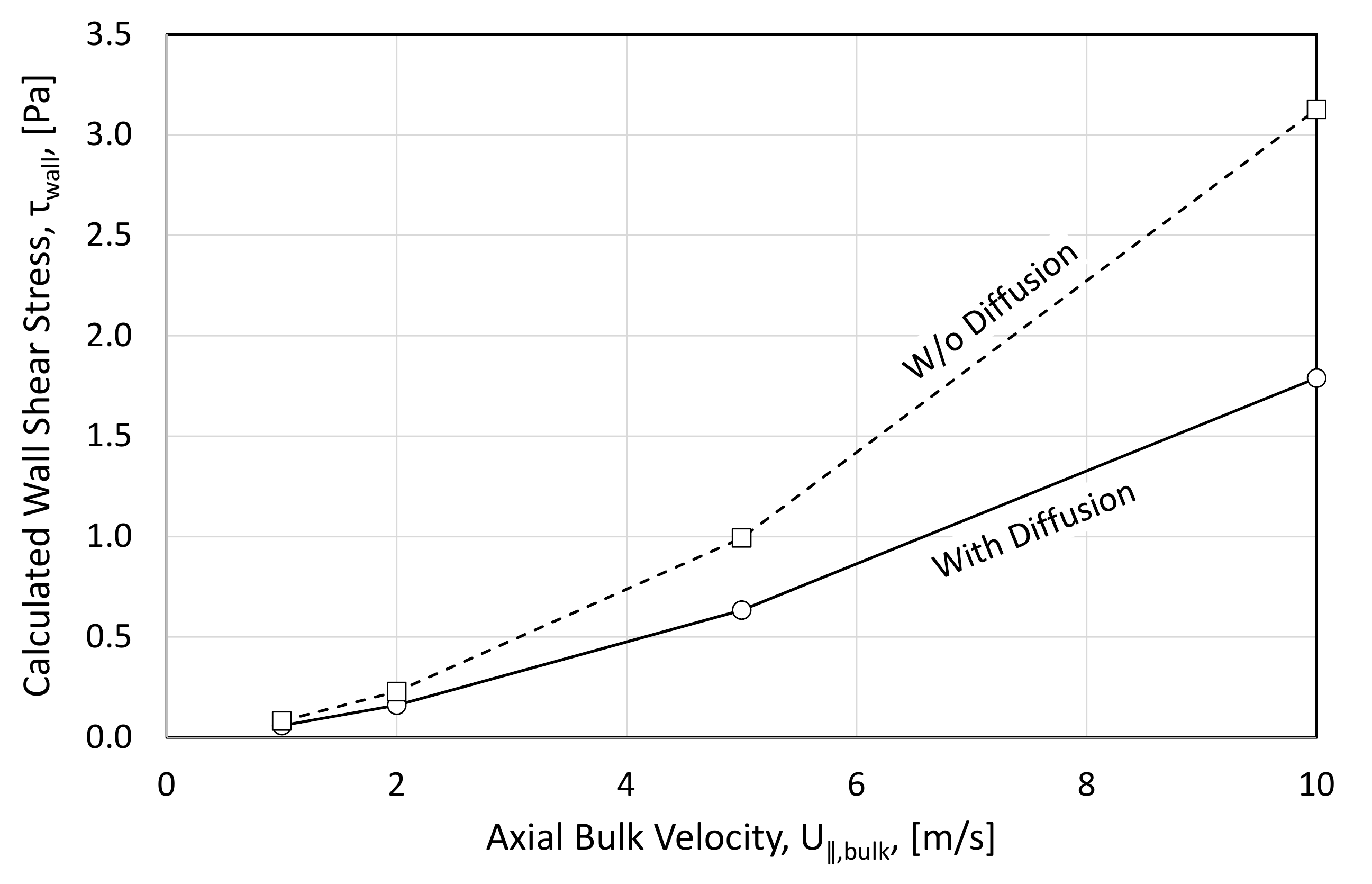}
	\caption{Calculated wall shear stress, $\tau_{wall}$, plotted against bulk flow velocity.}
	\label{fig:TauWall}  
	\end{center}
\end{figure}

\section*{\scalebox{.935}[1.0]{CONCLUSION}}
Employing mathematical modelling, it has been shown that the combined turbulent, diffusiophoretic and thermophoretic diffusion can have a siginficant effect on composition profiles in the turbulent boundary-layer for inert, multicomponent fluids.
This is of importance for the interpretation of rheology measurements to establish e.g. viscosity and thermal conductivity, since the fluid composition at the wall may differ significantly from the bulk composition.

Mathematical proof was given to support the following statements for inert mixtures:
\vspace{-3mm}
\begin{itemize}
	\setlength{\itemsep}{1pt}
  	\setlength{\parskip}{0pt}
  	\setlength{\parsep}{0pt}
	\item In the absence of supplementary scalar field gradients: 
	\begin{itemize}
		\setlength{\itemsep}{1pt}
  		\setlength{\parskip}{0pt}
  		\setlength{\parsep}{0pt}
		\item Non-constant composition profiles requires that $-\nicefrac{\nu_t}{Sc_t}$ is an eigenvalue of the matrix product $\bVec{D}\bVec{\Gamma}\bVec{\Lambda}$.
		\item Non-zero compositional gradients at the wall requires that $\det{\left(\bVec{\Gamma}\right)}=0$.
		\item Ideal mixture are not permitted to have non-zero compositional gradients at the wall.
	\end{itemize}
	\item In the presence of a temperature gradient:
	\begin{itemize}
		\setlength{\itemsep}{1pt}
  		\setlength{\parskip}{0pt}
  		\setlength{\parsep}{0pt}
		\item Non-zero compositional gradients are required to counter the thermophoresis.
		\item For ideal mixtures, the thermophoretic driving force coefficients must obey $\sum_{j=1}^N{z_jd_{T,j}}=0$.
	\end{itemize}
\end{itemize}





\section*{\scalebox{.935}[1.0]{ACKNOWLEDGEMENTS}}
Gratitude goes to all the colleagues at the research group of flow technology at dept. Process technology, SINTEF Industry, in Trondheim, Norway.
Without the vibrant research environment and fruitful discussions, this paper would not be.


\renewcommand{\nomname}{NOMENCLATURE}
\printnomenclature
\balance

\let\OLDthebibliography\thebibliography
\renewcommand\thebibliography[1]{
  \OLDthebibliography{#1}
  \setlength{\parskip}{2pt}
  \setlength{\itemsep}{1pt plus 0.3ex}
}
\def\bibindent{0cm}
\makeatletter
\renewcommand\@biblabel[1]{#1.}
\makeatother
\renewcommand{\refname}{\large \textnormal{REFERENCES}} 
\large{\bibliographystyle{NRC_test}}
\bibliography{References.bib}
\balance

\newpage
\section*{\scalebox{.935}[1.0]{APPENDIX - Dimensionless Variables}}
The model equations presented in this paper are presented in dimensionless form.
Dimensionless variables are denoted by superscript $+$. 
When making the conservation equations dimensionless, typical wall unit scaling is employed.
Selected scaled variables are given below. 

The shear velocity is defined as
\begin{equation}
	u_\tau=\sqrt{\nicefrac{\tau_w}{\rho_{f,wall}}}~,
\end{equation}
the dimensionless wall distance is defined as
\begin{equation}
	y^+=\nicefrac{yu_\tau}{\nu_{f,wall}}~,
\end{equation}
where $y$ is the normal distance to the wall.
$\nu_{f,wall}=\nicefrac{\mu_{f,wall}}{\rho_{f,wall}}$ is the kinematic viscosity at the wall, the dimensionless fluid velocity is defined as
\begin{equation}
	u_f^+=\nicefrac{u_f}{u_\tau}~,
\end{equation}
and the dimensionless mass flux is given by
\begin{equation}
	j^+=\nicefrac{j}{\rho_{f,wall}u_\tau}~.
\end{equation}

Fluid properties are typically converted to wall units by scaling with the value at the wall; e.g. 
\begin{align}
	&\rho_f^+=\nicefrac{\rho_f}{\rho_{f,wall}}~,\\
	&\mu_f^+=\nicefrac{\mu_f}{\mu_{f,wall}}~,\\
	&k_f^+=\nicefrac{k_f}{k_{f,wall}}~.
\end{align}
The dimensionless, turbulent thermal conductivity is defined as
\begin{equation}
	k_t^+=\nu_t^+\rho_f^+c_P^+\left(\nicefrac{Pr_{wall}}{Pr_t}\right)~,
\end{equation}
and the dimensionless, turbulent kinematic viscosity is modelled as\cite{Johansen91}
\begin{equation}
\small
\nu _{t,f}^{+}=\begin{cases}
	\left(\nicefrac{y^+}{11.15}\right)^3&\text{for $y^+<3.0$},\\
	\left(\nicefrac{y^+}{11.4}\right)^2 - 0.049774&\text{for $3.0\le y^+\le 52.108$},\\
	0.4y^+&\text{for $52.108<y^+$}.
\end{cases}
\end{equation}

Diffusivities are scaled by the fluid kinematic viscosity at the wall, e.g.
\begin{equation}
	D_{ij}^+ = \nicefrac{D_{ij}}{\nu_{f,wall}}~.
\end{equation}

The dimensionless temperature is given by
\begin{equation}
	T^+=\nicefrac{Tu_\tau\rho_{f,wall}c_{P,f,wall}}{q_w} - T_{wall}^{0+}~,
\end{equation}
where
\begin{equation}
	T_{wall}^{0+} = \nicefrac{T_{wall}u_\tau\rho_{f,wall}c_{P,wall}}{q_w}~,
\end{equation}
and $q_w=-\left. k_f\partial_\bot T\right|_{wall}$ is the wall heat flux.

The Prandtl number is given by
\begin{equation}
	Pr=\nicefrac{c_P\mu_f}{k_f}~.
\end{equation}
Constant turbulent Prandtl and Schmidt numbers of $Pr_t=0.85$ and $Sc_t=0.7$, respectively, were employed.


\balance

\end{large} 
\end{document}